\documentclass[letterpaper,journal]{IEEEtran}
\usepackage{amsmath,amsfonts}
\usepackage{algorithmic}
\usepackage{algorithm}
\usepackage{array}
\usepackage[caption=false,font=footnotesize]{subfig} 
\usepackage{textcomp}
\usepackage{stfloats}
\usepackage{url}
\usepackage{verbatim}
\usepackage{graphicx}
\usepackage{cite}
\usepackage{xcolor}
\begin{document}

\title{Efficient Joint Trajectory Optimization and Resource Allocation for Secure UAV Networks: A Deep Reinforcement Learning Approach}

\author{Abdulmalik Alwarafy \IEEEmembership{(Senior Member, IEEE)}, Mohamed Hayajneh \IEEEmembership{(Member, IEEE)}, Mustafa Al Juboori, Abdulaziz Almarzooqi, Tahnoon Almazrouei, Robera Abera
, and Kaleab Nega
}

\markboth{Journal of \LaTeX\ Class Files,~Vol.~14, No.~8, August~2021}%
{Shell \MakeLowercase{\textit{et al.}}: A Sample Article Using IEEEtran.cls for IEEE Journals}


\maketitle

\begin{abstract}
Unmanned Aerial Vehicles (UAVs) are increasingly deployed in wireless networks; however, ensuring secure communication and efficient energy usage remains challenging. This paper addresses the joint trajectory optimization and resource allocation problem in UAV networks in the presence of a tethered UAV flying center and a malicious ground eavesdropper. In the proposed model, UAVs are equipped with energy harvesting capabilities and can adaptively switch between transmitting, jamming, and relaying roles to deliver data to a tethered UAV center for optimized network performance. An optimization problem is formulated to maximize a utility function that balances secrecy rate, energy consumption, and interference. To address the high non-convexity of the problem and its real-time decision requirements, a Proximal Policy Optimization (PPO)–based Deep Reinforcement Learning (DRL) algorithm is proposed to solve the problem and learn the network dynamics. Extensive simulation results demonstrate that the proposed approach achieves superior secrecy performance, lower energy consumption, and reduced interference compared to conventional and DRL-based methods.

\end{abstract}


\vspace{-0.35cm}
\section{INTRODUCTION} \label{Introduction}
\IEEEPARstart{U}{nmanned} 
Aerial Vehicles (UAVs) have become a fundamental component of modern wireless communication systems due to their flexible deployment and ability to rapidly provide coverage in environments lacking terrestrial infrastructure \cite{Zhao2025Survey, Aboueleneen2023Deep}. UAVs now support a wide range of applications, including environmental monitoring \cite{bai2023towards}, precision agriculture \cite{Seo2025Density}, disaster response \cite{Jankovic2025Disaster}, surveillance \cite{Nguyen2025Person}, and defense operations \cite{Freitas2025Multi}. As multi-UAV networks expand, effective coordination among UAVs becomes important to ensure reliable communication and robust mission performance.

Despite their capabilities, UAV-enabled wireless networks remain highly vulnerable to cyber-physical threats \cite{Poorvi2025Reliable, Lei2026Joint}. The open nature of wireless channels and UAV mobility expose them to eavesdropping, jamming, spoofing, and other adversarial attacks \cite{yu2023cybersecurity}. These threats jeopardize confidentiality and degrade communication quality, thereby making secrecy rate maximization a key requirement. Achieving strong secrecy performance requires careful coordination of UAV trajectories, transmission power, interference management, and energy usage, which is especially challenging given the limited battery capacity of UAV platforms. In addition, traditional optimization techniques often struggle with these intertwined challenges due to the high-dimensional, nonlinear, and rapidly dynamic characteristics of UAV networks \cite{Zhao2025Survey}. Furthermore, time-varying channel conditions, evolving battery states, and adaptive adversaries make real-time decision-making quite challenging. Deep Reinforcement Learning (DRL) has recently emerged as a powerful tool for addressing such problems and enabling UAVs to autonomously learn optimal mobility and resource-allocation strategies through continuous interaction with the wireless environment \cite{Alwarafy2022Frontiers, Aboueleneen2023Deep, Zhao2025Survey}.

Motivated by these challenges, this work introduces a novel dynamic-role UAV architecture in which each UAV can seamlessly switch among transmitting, jamming, and relaying  roles. Unlike traditional fixed-role UAV systems, dynamic role selection allows more efficient utilization of available UAVs and significantly enhances secrecy performance under varying channel conditions, threat levels, and energy constraints. To further address UAV energy limitations, we propose to deploy a tethered flying center, which is a specialized aerial node connected to a ground power source \cite{Belmekki2022Unleashing, lee2025d}. This node acts simultaneously as a continuous wireless charging station and a centralized coordination hub to enable uninterrupted recharging and an aggregation point for legitimate data transfer. Consequently, the primary objective of this paper is to jointly optimize 3D UAV trajectories, transmit power, harvested energy, and dynamic role selection to maximize secrecy rate while minimizing overall energy consumption and interference. This problem is inherently combinatorial and non-convex due to the high-dimensionality and coupled nature of the optimization variables. To address these challenges, we develop a Proximal Policy Optimization (PPO)-based DRL framework capable of learning adaptive control policies in real time, even under partial state information and adversarial mobility. The key contributions of this paper are summarized as follows:
\begin{itemize}
    \item We formulate a multi-objective, combinatorial, and non-convex optimization problem that jointly maximizes secrecy rate, minimizes energy consumption, and mitigates interference by optimizing UAV trajectories, transmit/jamming/relaying power, harvested energy, and dynamic-role UAV selection (transmit, jam, or relay).
    \item To address the computational complexity of the hybrid continuous–discrete decision space, we propose a centralized PPO-based DRL framework capable of learning mobility, power control, and mode-switching strategies. Unlike most prior works that rely on fixed UAV roles or perfect CSI, the proposed method operates under partial network knowledge and dynamically adapts to real-time network changes.
    \item We carefully design the DRL model parameters to incorporate UAV energy harvesting, tethered aerial charging, and dynamic role switching, and we show via simulations that the learned policy successfully optimizes the multi-objective utility function and captures network dynamics.
    \item Using extensive simulations, we demonstrate that the proposed PPO-based solution outperforms conventional optimization and DRL baselines in terms of secrecy performance, energy efficiency, convergence behavior, and interference reduction. Additionally, the model exhibits strong adaptability when deployed in online scenarios.
\end{itemize}

The remainder of the paper is organized as follows. Section \ref{Related_Work} reviews the relevant literature. Section \ref{System_Model_Problem_Formulation} discusses the proposed system model along with the problem formulation. Section \ref{sec3} outlines the PPO-based DRL framework, describes its main components, and summarizes the PPO algorithm. Section \ref{Performance_Evaluation} details the simulation setup and offers an in-depth analysis of the numerical results. Finally, Section \ref{Conculusion} provides concluding remarks.

\vspace{-0.35cm}
\section{Related Work} \label{Related_Work}
Physical-layer security in UAV-enabled wireless networks has attracted significant attention, particularly in scenarios involving mobile eavesdroppers and strict energy constraints. Many existing works improve secrecy performance by jointly optimizing UAV trajectory, transmit power, and jamming strategies. For example, in \cite{iqbal2023deep}, the authors maximized the secrecy rate by proposing a DRL approach to optimize UAV trajectory and beamforming, although power allocation is not considered. In \cite{jing2021maximizing}, a DRL-based security algorithm was proposed to learn optimal UAV flight strategies for secrecy rate enhancement. Similarly, the authors in \cite{shao2023secrecy} employed a PPO framework to jointly optimize UAV trajectory and transmit power under secrecy and operational constraints. The work in \cite{9161257} optimized both the positions and power levels of transmitting and jamming UAVs using DRL, yet energy consumption is not incorporated. Complementary research in \cite{8438310} deployed a mobile jamming UAV whose trajectory and jamming power are optimized to degrade eavesdropper reception, although energy efficiency remains unaddressed.

Relay-assisted secure UAV networks have also been explored for improving secrecy and coverage. In \cite{li2018secrecy}, secrecy performance is enhanced by jointly designing transmit beamforming and artificial noise at a relay UAV using a DRL approach. A PPO-based DRL method was introduced in \cite{10087301} for joint trajectory and power optimization in a relay–jammer UAV configuration. In \cite{9390913}, the authors proposed a DRL-based UAV agents to learn optimal flight strategies that maximize secrecy rates, while \cite{9122470} analyzed secrecy outage performance for UAVs via statistical modeling.

Energy-aware secure communication has received increasing attention due to the limited endurance of UAVs. For example, a full-duplex UAV model was proposed in \cite{9022999} to jointly optimize trajectory and power for enhancing secrecy and energy efficiency. A hybrid genetic–particle-swarm optimization method was proposed in \cite{9642178} to maximize secrecy throughput via joint power allocation, UAV positioning, and SWIPT parameters. Similarly, the authors in \cite{9713997} explored UAV deployment and jamming behavior using deep learning under various CSI assumptions, although energy harvesting constraints are not considered. Additional works such as \cite{10693783} and \cite{10118891} addressed secrecy–energy efficiency tradeoffs using convex approximation and DRL-based solutions, respectively, with an emphasis on trajectory and beamforming optimization.

Another important direction in the literature focuses on physical-layer security improvements using reconfigurable environments such as reconfigurable intelligent surfaces (RISs). While RISs are not part of our system architecture, these studies are relevant as they examine advanced secrecy-enhancing mechanisms. In \cite{yang2020deep}, DRL was used to coordinate UAV trajectory and RISs beamforming to enhance secrecy. The works in \cite{zhang2022distributed} and \cite{liu2023drl} optimized RISs phase shifts and transmit strategies to mitigate eavesdropping in multi-user settings.

More general UAV secrecy frameworks consider adversarial mobility and directional transmission. For example, the authors \cite{zhang2020deep} proposed a DRL-based method to investigate robust secrecy optimization under uncertain eavesdropper locations by leveraging UAV mobility and 3D directional transmission. Similarly, the work in \cite{wen2022securing} applied DRL to jointly optimize UAV trajectory, power allocation, and scheduling in the presence of a mobile attacker.

Despite these advances, the majority of prior works rely on static role assignments, where UAVs operate exclusively as transmitters, jammers, or relays and do not incorporate dynamic role switching based on instantaneous network conditions, including our previous works in \cite{aboueleneen2023secure, Alwarafy2025Deep}. Moreover, most studies assume access to fixed ground control centers or static charging stations, overlooking the potential benefits of airborne energy support. In contrast to these limitations, our work introduces, for the first time in the literature, a flexible UAV network in which each UAV can dynamically switch between transmitting, jamming, and relaying roles. Furthermore, we integrate a tethered UAV flying center that provides continuous wireless charging and centralized coordination. This combination of dynamic role adaptation, wireless energy provisioning, and PPO-based decision-making enables our system to jointly optimize secrecy rate, energy consumption, interference level, and UAV mobility in real time.

\textbf{NOTATIONS:} $\mathbf{A}$, $\boldsymbol{a}$, $a$, and $\mathcal{A}$ represent a matrix, vector, scalar, and set, respectively. The notation $Tr(a)$ refers to the matrix trace $a$, while $\text{diag}(\mathbf{a})$ represents a diagonal matrix with the vector $\mathbf{a}$ as its diagonal. The superscripts $(\cdot)^H$ and $(\cdot)^T$ refer to the Hermitian and transpose operations, respectively. $\left| a \right|$ and $\left| \mathbf{a} \right|$ denote the magnitude of a complex number and the Euclidean norm of a vector, respectively. $\mathbb{C}^{x \times y}$ means the space of $x \times y$ complex matrices, and $\bigotimes$ denotes the outer product of two vectors.

\vspace{-0.35cm}
\section{System Model and Problem Formulation} \label{System_Model_Problem_Formulation}
This section provides an explanation of the proposed system model and formulates the corresponding optimization problem.

\begin{figure}[ht!]
	\centering
	\includegraphics[width=9 cm, height=6.5cm]{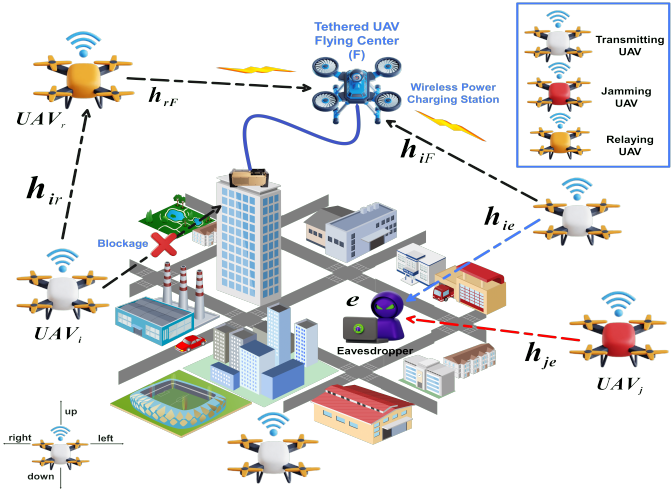}
	\caption{Illustration of our proposed UAV system model.}
	\label{Fig1}
\vspace{-6.mm}
\end{figure} 

\vspace{-0.35cm}
\subsection{Network Model} \label{NETWORK_MODEL}
The overall network architecture is illustrated in Fig. 1. The system consists of multiple UAVs operating within a defined coverage area to collect data and forward it to a centrally positioned tethered UAV flying center (F) \cite{Saif2021Energy, Belmekki2022Unleashing}, which serves as the network controller responsible for data aggregation and processing. Each UAV operates in a dynamic-role manner \cite{Gao2024Dynamic}, which enables it to function as a transmitter, jammer, or relay as needed. In the transmitting mode, a UAV sends data directly to the flying center whenever a line-of-sight (LoS) link is available. If a direct LoS connection is absent, the UAV switches to the relaying mode to forward data from the transmitting UAV to the flying center to maintain reliable communication. The network also assumes the presence of an unauthorized passive ground eavesdropper attempting to intercept the uplink communications between the UAVs and the tethered UAV flying center (F) \cite{Pandey2022Security}. To enhance physical layer security, it is assumed that UAVs can also operate in the jamming mode where they emit structured artificial noise along the null space of the legitimate UAV-to-F communication channel \cite{Wang2022Physical, aboueleneen2023secure, Alwarafy2025Deep}. This jamming strategy ensures that the confidential transmitted information remains secure. Since the tethered UAV flying center (F) possesses prior knowledge of the channel state information (CSI) and the characteristics of the artificial noise, it can effectively suppress the noise and accurately recover the intended signals. Conversely, the eavesdropper lacking access to such information experiences significantly degraded reception quality, thereby obstructing its ability to decode the transmitted data. It is noteworthy that artificial noise transmitted by jamming UAVs can unintentionally interfere with legitimate UAV links to degrade their signal quality and reduce network performance \cite{Zhang2020UAVSecure, Pandey2022Security}. To account for this effect, we explicitly incorporate and constrain the resulting interference in the optimization problem \textbf{P1} by enforcing that it remains below a predefined threshold, as specified in constraint $C_3$. In our framework, the system is assumed to have knowledge of the jamming UAVs’ transmit power, which serves as a tractable and practical proxy for estimating interference. This assumption is realistic in coordinated UAV networks where a centralized controller or tethered UAV flying center (F) manages mission parameters and communication resources. Prior studies confirm that cooperative jamming UAVs commonly operate under centralized or semi-centralized control, where their transmit power levels are either pre-configured or dynamically adjusted by the controller \cite{Zhang2020UAVSecure}. Hence, it is reasonable to assume that the network has access to the jamming UAVs’ transmit power, particularly when these UAVs participate in an integrated physical-layer security mechanism within the same trusted infrastructure.

The tethered UAV flying center (F) illustrated in Fig. 1 is further assumed to be equipped with wireless power charging capabilities that enable it to replenish the energy of the transmitting, jamming, and relay UAVs, thereby sustaining their long-term operational performance \cite{Belmekki2022Unleashing}. In addition, it is assumed that all transmitting, jamming, and relay UAVs are equipped with energy-harvesting modules that convert the received radiated energy from the tethered UAV flying center (F) into usable power.

The network operates in discrete time slots $t$, during which the system state remains constant and then transitions to a new state in the subsequent slot. At each time slot, the network jointly determines the 3D locations of all UAVs; the transmission, jamming, and relaying power levels; the energy transferred from the tethered UAV flying center (F) to the transmitting, jamming, and relay UAVs; the role of each UAV; and the 3D position of the tethered flying center.

Let $N_t$, $N_j$, $N_r$, and $N_e$ denote the number of transmitting UAVs, jamming UAVs, relay UAVs, and unauthorized ground eavesdroppers, respectively. Denote $h_{ij}$ as the channel gain between the $i$th transmitting UAV and the $j$th jamming UAV; $h_{ir}$ as the channel gain between the $i$th transmitting UAV and the $r$th relay UAV; $h_{rj}$ as the channel gain between the $r$th relay UAV and the $j$th jamming UAV; $h_{ie}$ as the channel gain between the $i$th transmitting UAV and the $e$th eavesdropper; $h_{re}$ as the channel gain between the $r$th relay UAV and the $e$th eavesdropper; and $h_{je}$ as the channel gain between the $j$th jamming UAV and the $e$th eavesdropper. Furthermore, let $p_i$, $p_j$, and $p_r$ denote the transmit power of the $i$th transmitting UAV, the $j$th jamming UAV, and the $r$th relay UAV, respectively. Throughout this paper, the index $i$ refers to transmitting UAVs $\forall i \in N_i$, $j$ refers to jamming UAVs $\forall j \in N_j$, $r$ refers to relay UAVs $\forall r \in N_r$, and $e$ refers to ground eavesdroppers $\forall e \in N_e$.

\vspace{-0.35cm}
\subsection{Secrecy Communication Model}
This paper focuses on the uplink communication between the transmitting or relay UAVs and the tethered UAV flying center (F). Based on a probabilistic channel model, the communication link between the UAVs and the tethered UAV flying center (F) may experience either LoS or NLoS conditions \cite{aboueleneen2023secure, perera2017simultaneous, yao2019qos, Alwarafy2025Deep}. The corresponding LoS probability ($p_{LoS}$) and NLoS probability ($p_{NLoS}$) are characterized using this probabilistic model as follows \cite{aboueleneen2023secure, perera2017simultaneous, yao2019qos, Alwarafy2025Deep}:
\begin{equation}
p_{LoS} =   \frac{1}{(1+\beta \exp(-a[\frac{180}{\pi }arctan(\frac{h}{d})-\beta])} 
\label{Eq1}
\end{equation}
where the constants $a$ and $\beta$ depend on the characteristics of the urban environment, $d$ represents the distance between the UAVs and the tethered UAV flying center (F), and $h$ denotes the UAVs’ hovering altitude. Consequently, the NLoS probability is obtained from $p_{LoS}$ using $p_{NLoS} = 1 - p_{LoS}$. The corresponding path loss expressions for both LoS and NLoS conditions are expressed as \cite{Alwarafy2025Deep, alwarafy2017path, Alwarafy2018Performance}: 
\begin{equation}
PL_{LoS}(d_{TXRX})=\lambda_{LoS} + 10 n_{LoS} \log_{10}(\frac{4\pi f_c d_{TXRX}}{c})
\label{Eq2}
\end{equation}
\begin{equation}
\footnotesize
PL_{NLoS}(d_{TXRX})=\lambda_{NLoS} + 10 n_{NLoS} \log_{10}(\frac{4\pi f_c d_{TXRX}}{c})
\label{Eq3}
\end{equation}
where $\lambda_{LoS}$ and $\lambda_{NLoS}$ represent environment-dependent constants, $d_{TXRX}$ denotes the separation distance between the transmitter and receiver UAVs, $c$ denotes the speed of light, and $f_c$ denotes the carrier frequency. The average path loss is then computed by \cite{aboueleneen2023secure, perera2017simultaneous, yao2019qos, Alwarafy2025Deep, yao2021wireless}:
\begin{equation}
\overline{PL}(d_{TXRX}) = p_{LoS} \times PL_{LoS}+p_{NLoS} \times PL_{NLoS}
\label{Eq4}
\end{equation}
It is worth noting that representing the channel as LoS or NLoS through probabilistic switching (i.e., a Bernoulli draw per time slot) is also a valid approach to capture instantaneous CSI variations. Nevertheless, we deliberately employ the averaged path loss model in (\ref{Eq4}), consistent with \cite{yao2021wireless, aboueleneen2023secure, Alwarafy2025Deep}, for two main reasons: 1) to simplify the underlying system dynamics for tractable optimization and stable DRL convergence, and 2) to account for partial LoS and NLoS propagation scenarios.

The channel gain for any transmitter–receiver communication link, as depicted in Fig. \ref{Fig1}, is given by \cite{RN282, alwarafy2022hierarchical, alwarafy2021DeepRAT, Alwarafy2025Deep}:
\begin{equation}
h_{TXRX}(t) = 10^{\frac{-\overline{PL}(d_{TXRX})}{10}} | \Gamma_{TXRX}(t) |^{2}, 
\label{Eq3_1}
\end{equation}
where $\Gamma_{TXRX}(t)$ denotes the small-scale fading component associated with the transmitter–receiver communication channel. Assuming an OFDM-based system operating under flat-fading conditions \cite{Alwarafy2025Deep}, the upper bound of the uplink data rate for all communication links in Fig. \ref{Fig1} at time slot $t$ is expressed as shown in (\ref{Eq5}). Hence, the transmission data rate from the $i$th transmitting or the $r$th relay UAVs to the tethered UAV flying center (F) at $t$ is \cite{alwarafy2021DeepRAT, Alwarafy2025Deep, alwarafy2022hierarchical, aboueleneen2023secure}:
\begin{equation}
\footnotesize
R_{i/rF} (t) = \omega(t) \log_2 \Big(1 + \frac{p_{i/rF}(t) h_{i/rF}(t)}{\omega(t) N_0} \Big), \forall i, r \in N_t, N_r
\label{Eq5}
\end{equation}
where $\omega(t)$ denotes the allocated bandwidth (which is considered fixed and assigned in advance within our system model), $p_{i/rF}(t)$ denotes the transmit power of the $i$th/$r$th transmitting/relaying UAV and the tethered UAV flying center (F), $h_{i/rF}(t)$  (as defined in (\ref{Eq3_1})) represents the channel gain between the $i$th/$r$th transmitting/relaying UAV and the tethered UAV flying center (F) at $t$, and $N_0$ represents the noise power spectral density. The data rate corresponding to the transmission from the $i$th/$r$th transmitting/relaying UAV to the $e$th eavesdropper at $t$ is:
\begin{equation}
\begin{array}{l}
R_{i/re} (t) = \omega(t) \log_2 \Big(1 + \frac{p_{i/r}(t) h_{i/re}(t)}{\omega(t) N^{\prime}_0 + p_j(t) h_{je}(t)} \Big), \\
\forall i, r, j \in N_t, N_r, N_j
\label{Eq6}
\end{array}
\end{equation}
where $h_{i/re}(t)$ represents the channel gain between the $i$th/$r$th transmitting/relaying UAV and the $e$th eavesdropper, $h_{je}(t)$ represents the channel gain between the $j$th jamming UAV and the $e$th eavesdropper, as illustrated in Fig. \ref{Fig1}, and $p_j(t)$ denotes the artificial noise power transmitted by the $j$th jamming UAV, which is intentionally employed to degrade the eavesdropper’s ability to intercept legitimate communications.

To ensure secure communication between transmitting and relaying UAVs and the tethered UAV flying center (F), it is essential to maximize the secrecy rate of the legitimate channels. The average secrecy rate of the transmitting and relaying UAVs is computed as the difference between the data rate sent from the $i$th/$r$th UAV to the tethered UAV flying center (F) and the data rate associated with the $i$th/$r$th UAV’s transmission to the $e$th eavesdropper, i.e. \cite{ Alwarafy2025Deep,aboueleneen2023secure}: 
\begin{equation}
\begin{aligned}
R_{i}^{sec}(t) = R_{iF}(t) - R_{ie}(t), ~~~ \forall i, e \in N_t, N_e \\
R_{r}^{sec}(t) = R_{rF}(t) - R_{re}(t), ~~~ \forall r, e \in N_r, N_e
\end{aligned}
\label{Eq7}
\end{equation}

Hence, the total secrecy rate achieved by all transmitting and relying UAVs in the UAV network shown in Fig. \ref{Fig1} is:
\begin{equation}
\footnotesize
\begin{aligned}
R_{tot}^{sec}(t) = \sum_{i=1}^{N_t} R_{i}^{sec}(t) + \sum_{r=1}^{N_r} R_{r}^{sec}(t) = ~~~~ \\ \sum_{i=1}^{N_t} R_{iF}(t) - R_{ie}(t) + \sum_{r=1}^{N_r} R_{rF}(t) - R_{re}(t),
\end{aligned}
\label{Eq_totalSecrecyRate}
\end{equation}

\vspace{-0.4cm}
\subsection{Energy Model}
This subsection describes the energy model of the proposed system illustrated in Fig. \ref{Fig1}. Three energy components are considered: the energy consumed by the transmitting or relaying UAVs for data transmission, the energy expended by the jamming UAVs, and the energy harvested by the UAVs.

\subsubsection{UAVs Data Transmission Energy}
The energy consumed by the $i$th/$r$th transmitting/relaying UAV to transmit data to the tethered UAV flying center (F) is \cite{Alwarafy2025Deep}:

\begin{equation}
e_{i/rF}^{tr}(t) = \frac{p_{i/rF}(t) l_{i/rF}(t)} {R_{i/rF}(t)} = \frac{p_{i/rF}(t) l_{i/rF}(t)} {\omega(t) \log_{2}\big( 1+\frac{p_{i/rF}(t) h_{i/rF}(t)}{\omega(t) N_0}\big)}
\label{Eq9}
\end{equation}
where $l_{i/rF}(t)$ denotes the data packet size transmitted from the $i$th/$r$th transmitting/relaying UAV to the tethered UAV flying center (F) at time slot $t$.

\subsubsection{UAVs Jamming Energy}
The primary role of the jamming UAVs shown in Fig. \ref{Fig1} is to hinder the ground eavesdropper from correctly decoding the information transmitted by the transmitting or relaying UAVs. This is achieved by reducing the eavesdropper’s received SINR below a threshold via controlled jamming \cite{Alwarafy2025Deep}. The energy consumed by the $j$th jamming UAV to transmit artificial noise and interfere with signal received by the $e$th eavesdropper is \cite{aboueleneen2023secure, Alwarafy2025Deep}:
\begin{equation}
\footnotesize
 e_{je}^{tr}(t)=\frac{p_{je}(t) l_{je}(t)}{R_{i/re}(t)}=\frac{p_{je}(t) l_{je}(t)}{\omega(t) \log_{2}\big( 1+\frac{p_{i/re}(t) h_{i/re}(t)}{\omega(t) N^{\prime}_0 + p_{je}(t) h_{je}(t)} \big)} 
\label{Equation:10}
\end{equation}
where $l_{je}(t)$ denotes the data packet size transmitted from the $j$th jamming UAV to the $e$th ground eavesdropper at time slot $t$. In our considered model, jamming UAVs are assumed to transmit artificial noise simultaneously with transmitting/relaying UAVs within the same time slots. Nevertheless, the energy consumption defined in (\ref{Equation:10}) is evaluated based on the effective jamming duration required to introduce adequate interference at the eavesdropper \cite{Alwarafy2025Deep, Li2019UAV, aboueleneen2023secure}. This modeling assumption captures the inherent trade-off between jamming effectiveness and energy consumption under practical transmission conditions. More specifically, it reflects an energy-aware security design in which jamming is activated only to the minimum extent necessary to guarantee secure communication, thereby avoiding unnecessary energy consumption \cite{Alwarafy2025Deep}. The effective jamming duration corresponds to the minimum time needed to degrade the eavesdropper’s received signal quality below a target secrecy threshold. This duration is implicitly influenced by key parameters such as the jamming transmit power, CSI, and the eavesdropper’s signal detection and decoding capability \cite{Zhang2020UAVSecure}.

\subsubsection{UAVs Harvesting Energy}
In the system model shown in Fig. \ref{Fig1}, wireless radio-frequency (RF) energy harvesting is employed to recharge all UAVs in the network to extend their operational lifetime \cite{yao2021wireless, lu2018wireless, 9405902, 10190121}. It is assumed that the tethered UAV flying center (F) transmits RF signals that act as a controllable energy source to provide continuous powering of all UAVs for enhanced quality of service (QoS) \cite{9405902, yang2019energy}. It is assumed all UAVs are equipped with RF energy harvesting circuitry capable of converting the received power from the tethered UAV flying center (F) into usable energy, similar to \cite{Alwarafy2025Deep, aboueleneen2023secure, 9405902, yang2019energy}.

In this paper, energy-harvesting management is incorporated as a key operational component. The harvested energy is primarily influenced by two factors: the path loss between the tethered UAV flying center (F) and the UAVs, and the amount of energy transmitted by the tethered UAV flying center (F) \cite{8214104, perera2017simultaneous, Alwarafy2025Deep}. Under a linear energy-harvesting model \cite{yao2021wireless}, the energy harvested by the $i$th/$j$th/$r$th UAV at $t$ is \cite{yao2021wireless, Alwarafy2025Deep}:
\begin{equation}
E^{h}_{i/j/r}(t) = h_{Fi/j/r}(t) \times \zeta_{i/j/r} \times e^{h}_{i/j/r}(t),
\label{Eq_harvested_energy}
\end{equation}
where $\zeta_{i/j/r}$ denotes the energy harvesting efficiency of $i$th/$j$th/$r$th UAV, $e^{h}_{i/j/r}(t)$ is the energy transferred from the tethered UAV flying center (F), and $h_{Fi/j/r}(t)$ is the channel gain for wireless energy transfer between the tethered UAV flying center (F) and the $i$th/$j$th/$r$th UAV, which is modeled as \cite{alwarafy2021DeepRAT} $h_{Fi/j/r}(t) = 10^{\frac{-\overline{PL}(d_{Fi/j/r})}{10}} \left| \Gamma_{Fi/j/r}(t) \right|^{2},$
where $\overline{PL}(d_{Fi/j/r})$, $d_{Fi/j/r}$, and $\Gamma_{Fi/j/r}(t)$ represent the average path loss, the distance, and small-scale fading component between the tethered UAV flying center (F) and the $i$th/$j$th/$r$th UAV, respectively.

The flight energy consumption of the $i$th transmitting UAV, the $r$th relaying UAV, and the $j$th jamming UAV accounts for several energy components, including hovering, mean rotor-induced, propulsion, and blade profile energies \cite{Alwarafy2025Deep, jing2021maximizing}. Consequently, the total flight energy consumption is \cite{Alwarafy2025Deep, jing2021maximizing, duo2020energy, yan2024energy}:
\begin{equation}
e_{i/j/r}^f(t)=\frac{P_f \left\| \boldsymbol{q}_{i/j/r}(t+1)-\boldsymbol{q}_{i/j/r}(t) \right\|} {v},
\label{Eq_flight_energy_consumption}
\end{equation}
where $P_f$ denotes UAVs' flight power, $v$ represents UAVs' velocity, and $\left\| \boldsymbol{q}_{i/j/r}(t) \right\|$ specifies the 3D position of the $i$th/$j$th/$r$th transmitting/jamming/relaying UAV at time slot $t$, where $\boldsymbol{q}_{i/j/r}(t) = [x_{i/j/r}(t),\, y_{i/j/r}(t),\, z_{i/j/r}(t)]$. The UAVs’ 3D positions $\left\| \boldsymbol{q}_{i/j/r}(t) \right\|$ are subject to the following constraints:
\begin{equation}
\left\| \boldsymbol{q}_{i/j/r}(t+1)-\boldsymbol{q}_{i/j/r}(t) \right\| ^{2} \leq L^{2},
\label{Equation:14}
\end{equation}
\begin{equation}
\left\|\boldsymbol{q}_{i/j/r}(t)-\boldsymbol{q}_{i/j/r}(t) \right\|^{2} \geq  d_{min}^{2},
\label{Equation:15}
\end{equation}
where \(L = \tau v_{max}\) denotes the maximum 3D Euclidean distance a UAV can travel within a single time slot, where \(v_{max}\) is the UAV's maximum velocity and \(\tau\) is the duration of each time slot. In addition, \(d_{\text{min}}\) specifies the minimum separation distance that must be maintained at all times between any transmitting, jamming, and relaying UAVs.

All UAVs in Fig. \ref{Fig1} are assumed to be capable of wirelessly recharging their batteries using energy supplied by the tethered UAV flying center (F). The battery state of the $i$th transmitting UAV, $j$th jamming UAV, and $r$th relay UAV is given by $b_{i/j/r}(t) = [b_1(t), b_2(t), \dots, b_{N_t}(t)/b_{N_r}(t)/b_{N_j}(t)]$, where $b(t) \in [0, B_{\max}]$ represents the battery level at the start of time $t$, and $B_{\max}$ is the maximum UAV battery capacity. The battery energy at each time $t$ depends on the harvested energy, the current level of battery, and the energy consumed for data transmission, jamming, or relaying. To compute the battery state at the next slot, two cases are considered \cite{Alwarafy2025Deep, aboueleneen2023secure}:
\begin{itemize}
    \item If the $i$th/$j$th/$r$th transmitting/jamming/relay UAV is fully charged to $B_{\max}$, its battery energy at time $t$ becomes $B_{\max} - e_{i/j/r}^{tr}(t)$.
    \item If the UAV’s battery is not fully charged, then the battery level for the $i$th/$j$th/$r$th transmitting/jamming/relay UAV is updated as $E_{i/j/r}^{H}(t) + b_{i/j/r}(t) - e_{i/j/r}^{tr}(t)$.
\end{itemize}

Hence, the battery level of the $i$th/$j$th/$r$th transmitting/jamming/relay UAV at the next time slot ($t+1$) is expressed as \cite{Alwarafy2025Deep, aboueleneen2023secure}: $b_{i/j/r}(t+1) = \min \left\{ E_{i/j/r}^{h}(t)+ b_{i/j/r}(t) - e_{i/j/r}^{tr}(t), B_{max} \right\}.$

The total energy consumption of the network incorporates all previously defined energy components, and the total network energy usage is formulated as \cite{Alwarafy2025Deep, aboueleneen2023secure}:

\begin{equation}
\footnotesize
\begin{aligned}
E_{tot}(t) = \sum_{i=1}^{N_t} \Big( e^{tr}_{iF}(t) + e^f_i(t) + e_i^{h}(t) \Big) + ~~~~~~~~~~~~~~~~~~~\\
\sum_{j=1}^{N_j} \Big( e^{tr}_{je}(t) + e^f_j(t) + e_j^{h}(t) \Big) +
\sum_{r=1}^{N_r} \Big( e^{tr}_{rF}(t) + e^f_r(t) + e_r^{h}(t) \Big),
\end{aligned}
\label{Eq_totalEnergyEffeciency}
\end{equation}

\vspace{-0.5cm}
\subsection{Interference Model}
The total network interference corresponds to the aggregate interference generated by the jamming UAVs toward all legitimate transmitting and relay UAVs \((I_{ji} + I_{jr})\), as illustrated in Fig. \ref{Fig1}. This interference is given by:
\begin{equation}
I_{tot}(t) = \sum_{j=1}^{N_j} \sum_{i=1}^{N_i} I_{ji}(t) + \sum_{j=1}^{N_j} \sum_{r=1}^{N_r} I_{jr}(t),
\label{Eq_totalInterference}
\end{equation}
where $I_{ji}(t)$ and $I_{jr}(t)$ are the interference generated by the $j$th jammming UAV towards the $i$th transmitting UAV and $r$th relay UAV, respectively, at time slot $t$. 

\vspace{-0.35cm}
\subsection{Mobility Model} \label{mobility_model}
To capture realistic and time-varying network behavior, the movement of both aerial and ground nodes is taken into account. The eavesdropper is modeled using the Random Waypoint mobility pattern \cite{Zhong2023On}, in which it travels among randomly chosen points within the service region and pauses for a stochastic duration at each point before proceeding to another. Meanwhile, all UAVs in the network (transmitting, jamming, and relaying UAVs) adjust their 3D trajectories dynamically in discrete time slots under the coordination of the tethered UAV flying center (F). The flying center itself is assumed to move slowly within the operational zone while remaining tethered to a ground-based power source \cite{Saif2021Energy, Belmekki2022Unleashing} to guarantee continuous energy availability and stable coordination links.

\vspace{-0.45cm}
\subsection{Problem Formulation} \label{Problem_Formulation}
The objective of this work is to jointly optimize the UAVs’ 3D trajectories, transmission power, harvested energy, UAVs’ operational roles, and the 3D trajectory of the tethered UAV flying center in order to maximize the overall network multi-objective utility function, expressed in terms of secrecy rate, energy consumption, and interference. Unlike traditional approaches that fix the number of transmitting, jamming, and relaying  UAVs, the proposed model maintains a fixed total number of UAVs $N$, where each UAV can dynamically switch between transmitting, jamming, or relaying roles to best serve the current network state. The optimization problem is formulated as a multi-objective utility maximization problem that seeks to balance the following three interdependent performance goals:

\begin{itemize}
\item Secrecy rate: Maximize the total secrecy rate of the network by jointly boosting the data throughput of legitimate links while suppressing the portion potentially accessible to eavesdroppers, as defined in (\ref{Eq_totalSecrecyRate}).
\item Energy consumption: Reduce the UAVs’ total energy consumptions, including communication, flight, and charging components, as formulated in (\ref{Eq_totalEnergyEffeciency}).
\item Interference mitigation: Limit the interference generated by jamming UAVs on authorized receivers, as in (\ref{Eq_totalInterference}).
\end{itemize}

\begin{table*}[ht!]
\centering
\begin{equation}
\label{OPB}
\resizebox{\textwidth}{!}{$\displaystyle\begin{aligned}
&\textbf{P1:} \quad \underset{\mathbf{Q}, \boldsymbol{p}, \boldsymbol{e}^{h}, \boldsymbol{N}, \boldsymbol{S}, \boldsymbol{q}_{F}} {\text{maximize}} \sum_{t=1}^{T} U(t) = \sum_{t=1}^{T} \bigg( \underbrace{w_1 R_{tot}^{sec}(t)}_{\text{Total secrecy rate}} - \underbrace{w_2 E_{tot}(t)}_{\text{Total energy consumption}} - \underbrace{w_3 I_{tot}(t)}_{\text{Total interference}} \bigg) \quad \\
&\text{s.t.} \quad\\
&C_1:  R_{i/r}^{sec}(t) \geq R_{min}^{sec}, \quad \forall i \in N_t, \quad \forall r \in N_r \\
&C_2: \Gamma_{i/re}(t) < \Gamma_e^{min}, \quad \forall i \in N_t, \quad \forall r \in N_r, \quad \forall e \in N_e \\
&C_3: I_{ji/r}(t) < I_{i/r}^{max}, \quad \forall i \in N_t, \quad \forall r \in N_r, \quad \forall j \in N_j \\
&C_4:  b_{i/j/r}(t+1) = \min\Big( b_{i/j/r}(t) + E_{i/j/r}^{h}(t) - e^t_{i/j/r}(t) - e^f_{i/j/r}(t), B^{max} \Big), \quad \forall i \in N_t, \quad \forall j \in N_j, \quad \forall r \in N_r   \\ 
&C_5:  b_{i/j/r}(t) + E_{i/j/r}^{h}(t) - e^t_{i/j/r}(t) - e^f_{i/j/r}(t) \geq 0, \quad \forall i \in N_t, \quad \forall j \in N_j, \quad \forall r \in N_r  \\
&C_6: \left\| \mathbf{q}_{i/j/r}(t)-\mathbf{q}_{i/j/r}(t-1) \right\| ^{2} \leq L^{2}, \quad \forall i \in N_t, \quad \forall j \in N_j, \quad \forall r \in N_r   \\
&C_7: \left\|\mathbf{q}_u(t)-\mathbf{q}_v(t) \right\|^{2} \geq  d_{min}^{2}, \quad u,v \in (i, j, r), u \neq v, \quad \forall i \in N_t, \quad \forall j \in N_j, \quad \forall r \in N_r \\
&C_8:  x_{min} \leq x_{i/j/r}(t) \leq x_{max}, y_{min} \leq y_{i/j/r}(t) \leq y_{max}, z_{min} \leq z_{i/j/r}(t) \leq z_{max}, \quad \forall i \in N_t, \forall j \in N_j, \forall r \in N_r \\
&C_9: 0 \leq p_{i/j/r} (t) \leq P^{max},  \quad \forall i \in N_t, \quad \forall j \in N_j, \quad \forall r \in N_r \\
&C_{10}: 1 \leq N_t(t) \leq N, \quad 1 \leq N_j(t) \leq N, \quad 1 \leq N_r(t) \leq N  \\
&C_{11}: \boldsymbol{N}=[ N_t(t), N_j(t), N_r(t) ], \quad N_t(t) + N_j(t) + N_r(t) = N \\
&C_{12}: \boldsymbol{S}(t) = [ \mathcal{S}_0(t), \mathcal{S}_1(t), \ldots, \mathcal{S}_n(t), \ldots, \mathcal{S}_N(t) ], \quad \mathcal{S}_n(t) \in \{ 0, 1, 2 \}  \\
&C_{13}: x_{\min} \leq x_F(t) \leq x_{\max},\quad y_{\min} \leq y_F(t) \leq y_{\max},\quad z_{\min} \leq z_F(t) \leq z_{\max}, \quad\quad \boldsymbol{q}_F(t)=[ x_F(t), y_F(t), z_F(t) ].  \\
\end{aligned}$}
\end{equation}
\hrule
\vspace{-6.mm}
\end{table*}

Hence, the optimization problem can be expressed as \textbf{P1} in (\ref{OPB}), where $w_1, w_2,$ and $w_3$ in the utility function $U(t)$ are weighting factors for secrecy rate, energy consumption, and interference mitigation, respectively, satisfying $w_1 + w_2 + w_3 = 1.$ Problem \textbf{P1} is subject to the following constraints:
\begin{itemize}
    \item $C_1$ ensures that each transmitting or relaying UAV achieves a secrecy rate greater than or equal to \(R^{sec}_{min}\).
    \item $C_2$ is the jamming constraint, which ensures that jamming UAVs effectively disrupt the signals intercepted by eavesdroppers by reducing their received SNR below the minimum required sensitivity threshold $\Gamma_e^{min}$.
    \item $C_3$ is the interference control constraint, which guarantees that the interference caused by jamming UAVs to transmitting UAVs remains below a defined threshold $I_{i/r}^{max}$.
    \item $C_4$ ensures that the battery levels of transmitting, jamming, and relaying UAVs at the next time slot are bounded by the harvested energy and the maximum UAV battery capacity $B_{max}$.
    \item $C_5$ enforces that the remaining energy of each UAV is always non-negative.
    \item $C_6$ is the mobility constraint, which ensures that the displacement of each UAV between consecutive time slots does not exceed \(L = \tau v_{max}\).
    \item $C_7$ is the collision avoidance constraint, which guarantees that the minimum Euclidean distance between any two UAVs remains greater than the safety distance $d_{min}$.
    \item $C_8$ is the spatial constraint, which ensures that all transmitting, jamming, and relaying UAVs remain within the predefined 3D operational region.
    \item $C_9$ is the power constraint, which guarantees that the transmit power of transmitting, jamming, and relaying UAVs does not exceed $P_{max}$.
    \item $C_{10}$--$C_{12}$ define the dynamic-role UAV constraints, which ensure that the total number of transmitting, jamming, and relaying UAVs at time slot $t$ does not exceed the total number of UAVs in the network. In addition, these constraints specify the operational role of each UAV at time slot $t$, where $\mathcal{S}_n(t)=0,1,2$ correspond to transmitting, jamming, and relaying modes, respectively.
    \item $C_{13}$ ensures that the 3D coordinates of the tethered UAV flying center (F) remain within the designated operational region of the UAV network.
\end{itemize}

\vspace{-0.35cm}
\subsection{Analysis of Problem P1}\label{Why_DRL}
Problem \textbf{P1} is a highly challenging mixed-integer non-convex and combinatorial optimization problem due to the non-concave structure of its weighted multi-objective utility function, the secrecy-related constraints $C_1$ and $C_2$, and the discrete and combinatorial role-selection constraint $C_{12}$. In particular, the role assignment vector $\boldsymbol{S}(t)$ introduces integer decision variables that are tightly coupled with continuous control variables such as UAV positions $\mathbf{Q}(t)$, transmit powers $\boldsymbol{p}(t)$, and harvested energy $\boldsymbol{e}^{h}(t)$. Consequently, problem \textbf{P1} can be formally classified as a mixed-integer nonlinear programming (MINLP) problem with both combinatorial and continuous decision spaces \cite{floudas1995nonlinear, chiang2007layering}.

Although exhaustive search methods combined with classical optimization techniques \cite{Huang2020Reconfigurable} could theoretically yield globally optimal solutions, such approaches rapidly become infeasible as the solution space grows exponentially \cite{boyd2004convex}. This exponential growth is driven by the high dimensionality of the network, the large-scale UAV configuration, the continuous mobility, and power control variables, and the combinatorial explosion induced by dynamic role switching, where the number of possible role assignments scales $3^N$ per time slot, as illustrated in Fig. \ref{Fig1}. As a result, conventional analytical and numerical optimization techniques are ill-suited for solving \textbf{P1} in real time \cite{Alwarafy2022Frontiers}.

Compared with prior formulations such as \cite{aboueleneen2023secure, Alwarafy2025Deep}, the proposed problem \textbf{P1} extends the optimization scope by enhancing the objective function; explicitly incorporating transmitting, jamming, and relaying UAV decision variables; modeling the mobility of the ground eavesdropper; enabling dynamic role switching; and utilizing a tethered UAV flying center. While these additions substantially improve the realism and practical relevance of the system model, they further intensify the MINLP complexity. Therefore, a PPO-based DRL framework is adopted to solve \textbf{P1} in order to enable the flying center to efficiently learn near-optimal control policies and adapt to the network’s dynamic, time-varying, and combinatorial behavior, as detailed in Section \ref{sec3}.

\vspace{-0.35cm}
\section{PPO for Joint Trajectory Optimization and Resource Allocation in UAV Networks} 
\label{sec3}
\begin{figure}[t!]
	\centering
	\includegraphics[width=8cm, height=6cm]{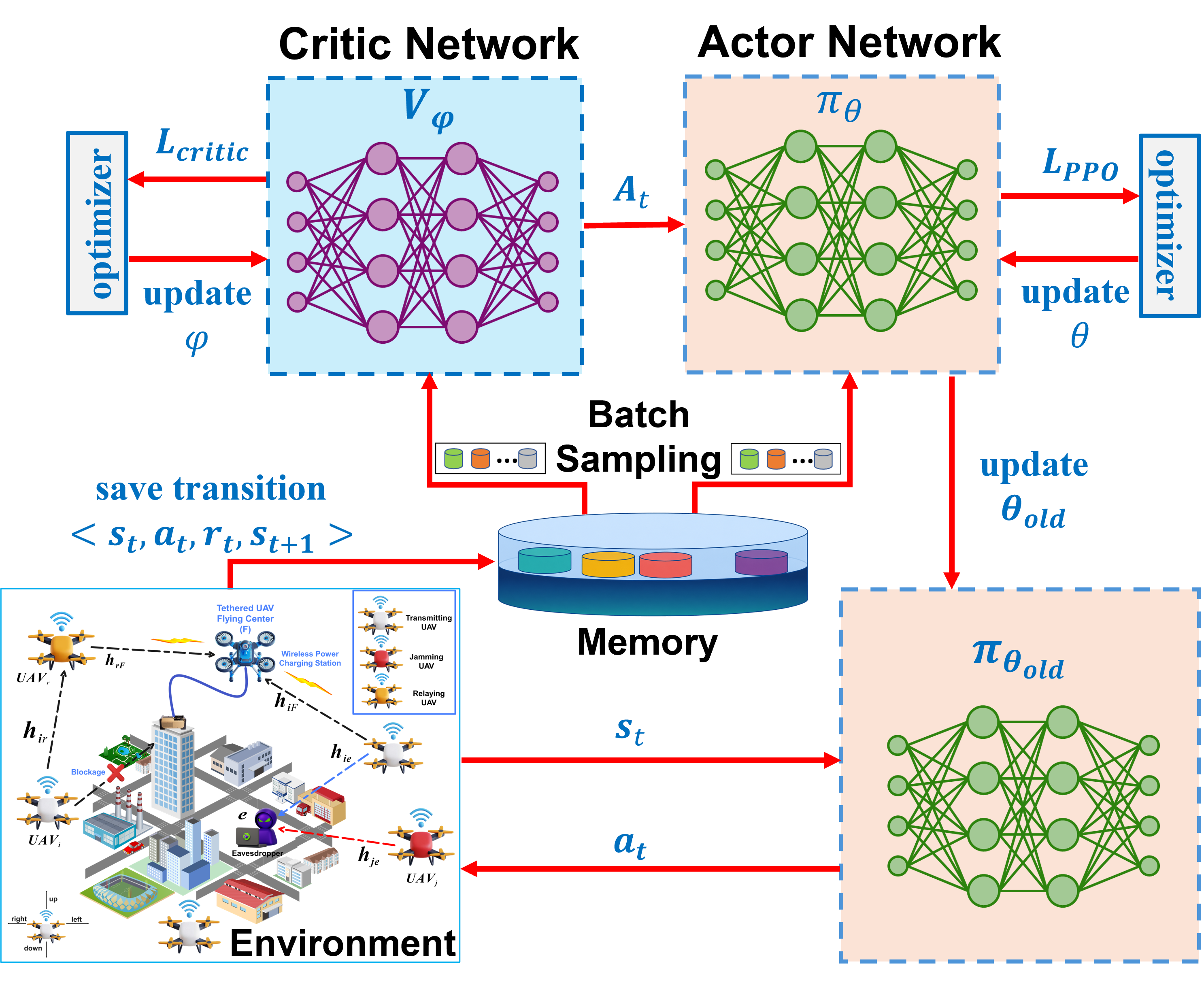}
	\caption{Training architecture of the proposed PPO model.}
	\label{PPO_architicture}      
\vspace{-6.mm}
\end{figure}

This section outlines the procedure for applying the DRL framework to address \textbf{P1} and to model the behavior of the UAV network in Fig. \ref{Fig1}. The optimization task is cast as a Markov decision process (MDP) described by the tuple ${\mathcal{S}, \mathcal{A}, p, r}$, where $\mathcal{S}$, $\mathcal{A}$, $p$, and $r$ denote the state set, action set, state-transition probability, and reward function, respectively \cite{Alwarafy2022Frontiers,Alwarafy2025Deep, alwarafy2022drl}. In this formulation, the UAV network represents the environment, whereas the tethered UAV flying center (F) acts as the learning agent that monitors the system state and selects actions accordingly. The environment then generates a reward based on the agent’s decision and transitions to the subsequent network state. The definitions of these components are provided below. The structure of the proposed PPO model designed to tackle \textbf{P1} is illustrated in Fig. \ref{PPO_architicture}.

\subsubsection{State Space} 
The state space is structured to include relevant information about the UAV network to help the agent (the tethered UAV flying center) learn an optimal policy. The flying agent is trained under the assumption that it only has access to partial knowledge about the UAV dynamics. Specifically, the agent is assumed to know only past performance data, such as achieved data rates and battery levels. It is important to note that the data rate equations from Section~(\ref{System_Model_Problem_Formulation}) still incorporate channel gain terms, which are essential for the agent's learning process. At time step $t$, the state space contains six key elements: the vector of downlink secrecy data rates from all transmitting and relaying UAVs to the flying UAV center at the previous time step, denoted by $\boldsymbol{R}^{sec}(t-1) \in \mathbb C^{(N_t + N_r) \times 1}$; the battery levels of all UAVs at the previous time step, represented by $\boldsymbol{b}(t-1) \in \mathbb C^{N \times 1}$; the energy consumed by all UAVs at the previous time step, denoted by $\boldsymbol{E}^{H}(t-1) \in \mathbb C^{N \times 1}$; the minimum secrecy rate requirements for the transmitting and relaying UAVs, represented by $R_{min}^{sec} \in \mathbb C^{1 \times 1}$, the vector of the roles of all UAVs in in the network at $t-1$, represented by $\boldsymbol{S}(t-1) \in \mathbb{C}^{N \times 1}$, and the vector of 3D coordinates of the tethered UAV flying center at $t-1$ represented by $\boldsymbol{q_F}(t-1) \in \mathbb{C}^{3 \times 1}$. Consequently, the state space is a vector of length $4N_t + 4N_r + 3N_j + 4$. The state space is:
\begin{equation}
\label{Eq_stateSpace}
\scriptsize
\begin{aligned}
\mathcal{S}(t)=\Big[&
\boldsymbol{R}^{\mathrm{sec}}(t-1)  \in \mathbb{C}^{(N_t+N_r)\times 1},
\boldsymbol{b}(t-1)  \in \mathbb{C}^{N\times 1},
\boldsymbol{E}^{H}(t-1)  \in \mathbb{C}^{N\times 1}, \\
& R_{\min}^{\mathrm{sec}} \in \mathbb{C}^{1\times 1}, 
\boldsymbol{S}(t-1)  \in \mathbb{C}^{N\times 1},
\boldsymbol{q}_{F}(t-1) \in \mathbb{C}^{3\times 1}
\Big],
\end{aligned}
\end{equation}

\subsubsection{Action Space}
In our system configuration illustrated in Fig. \ref{Fig1}, the decisions executed by the tethered UAV flying center (i.e., agent) are intended to optimize the objective function of \textbf{P1} through simultaneous adjustment of the system’s control variables. Accordingly, at each time step $t$, the agent selects the following actions: the matrix representing the 3D positions of all transmitting, jamming, and relaying UAVs, $\mathbf{Q}(t) \in \mathbb C^{N \times 3}$; the vector specifying the transmit, jamming, and relaying power levels for these UAVs, $\boldsymbol{p}(t) \in \mathbb C^{N \times 1}$; the vector describing the amount of energy supplied from the charging flaying station to all UAVs, $\boldsymbol{e}^h(t) \in \mathbb C^{N \times 1}$, the vector describing the number of UAVs acting as transmitting, jamming, and relaying at time step $t$, $\boldsymbol{N}(t) \in \mathbb C^{3 \times 1}$, the vector representing the roles of each UAV, $\boldsymbol{S}(t) \in \mathbb C^{N \times 1}$, and the vector of the 3D coordinates of the flying UAV center at $t$, denoted by $\boldsymbol{q}_{F}(t) \in \mathbb C^{3 \times 1}$. The resulting action vector has a total length of $6N+6$, with an action space of:
\begin{equation}
\label{Eq_actionSpace}
\scriptsize
\begin{aligned}
\mathcal{A}(t) = \Big[ \mathbf {Q}(t) \in \mathbb C^{ N \times 3},  \boldsymbol {p}(t) \in \mathbb C^{ N \times 1},  \boldsymbol {e}^h(t) \in \mathbb C^{ N \times 1}, \\
\boldsymbol{N}(t) \in \mathbb C^{3 \times 1}, \boldsymbol{S}(t) \in \mathbb C^{N \times 1}, 
\boldsymbol{q}_{F}(t) \in \mathbb C^{3 \times 1}\Big], 
\end{aligned}
\end{equation}

\subsubsection{Reward Function} 
In the optimization problem \textbf{P1}, the aim is to maximize a multi-objective utility function $U(t)$ that includes secrecy rate performance, energy considerations, and interference suppression. At each time step, the flying agent computes a reward by evaluating the observed state of the UAV network in (\ref{Eq_stateSpace}) together with the executed action in (\ref{Eq_actionSpace}). Accordingly, the instantaneous reward is:
\begin{equation}
r(t) = \sum_{t=1}^{T} \bigg( w_1 R_{tot}^{sec}(t) - w_2 E_{tot}(t) - w_3 I_{tot}(t) \bigg),
\label{Eq_rewardFunction}
\end{equation}

The overall utility maximization task can then be framed using this per-step reward. The expected cumulative discounted return over a time horizon $T$ is given by ${\cal R} = \mathbb{E} \left[ \sum_{t=1}^{T} \gamma^{t-1} r(t) \right],$ where $0 \leq \gamma \leq 1$ denotes the discount factor \cite{Alwarafy2022Frontiers}. The objective is to learn the optimal policy $\pi^{*}$ that prescribes the UAVs’ positions, transmission powers, harvested energy levels, UAVs' roles, and tethered flying UAV center, so as to maximize the long-term return. Formally, this is written as $\pi^{*} = \underset{\pi}{\text{argmax}} ~ {\cal R} = \underset{\pi}{\text{argmax}} \sum_{t=1}^{T} \gamma^{t-1} r(s(t), a(t))$.

\vspace{-0.35cm}
\subsection{The Proposed PPO Algorithm} 
As discussed previously in Subsection II.\ref{Why_DRL}, the optimization problem in \textbf{P1} exhibits considerable computational complexity that makes it impractical for conventional analytical or numerical techniques \cite{Alwarafy2022Frontiers}. To overcome these difficulties, we adopt the PPO algorithm to learn the optimal control strategy $\pi^{*}$, similar to \cite{Alwarafy2025Deep}. PPO has proven to be a reliable and effective DRL method, particularly well suited for scenarios involving large-scale and continuous state-action domains.

PPO operates as a policy gradient algorithm structured around the actor-critic paradigm \cite{schulman2017proximal, Alwarafy2025Deep}. It is essentially an advancement over the trust region policy optimization (TRPO) method developed to incorporate mechanisms to limit excessive policy updates. The algorithm employs importance sampling to measure the deviation between the updated and previous policies to provide controlled policy refinement while maintaining stability \cite{schulman2017proximal}. The optimal policy ($\pi^{*}$) is derived via training deep neural networks to minimize the following Bellman-style loss \cite{schulman2017proximal, Alwarafy2025Deep}:
\begin{equation}
L(\theta ) = {{\mathbb{E}}_t}\left[ {\log {\pi _\theta }\left( {{a_t}\mid {s_t}} \right){A_t}} \right],
\end{equation}
where $\theta$ denotes the parameters of the actor and critic networks, and $A_t$ represents an estimate of the advantage function at time step $t$, as shown in Fig. \ref{PPO_architicture}. In TRPO, the ratio between the updated policy and the previous one is expressed as $ \psi (\theta ) = \frac{{{\pi _\theta }\left( {{a_t}\mid {s_t}} \right)}}{{{\pi _{{\theta _{old}}}}\left( {{a_t}\mid {s_t}} \right)}}$. Using this ratio, an alternative policy objective can be written as \cite{schulman2017proximal, Alwarafy2025Deep}:
\begin{equation}
L_{TRPO}(\theta ) = {{\mathbb{E}}_t}\left[ {\psi (\theta ){A_t}} \right].
\end{equation}

Maximizing $L_{TRPO}(\theta)$ directly, however, may lead to excessively large adjustments to the policy. To prevent such destabilizing updates, PPO introduces a modified loss, $L_{PPO}(\theta)$, which constrains changes to the policy through a clipping operation. The PPO objective is defined as \cite{schulman2017proximal}:
\begin{equation}
\footnotesize
L_{PPO}(\theta ) = {{\mathbb{E}}_t}\left[ {\min \left( {\psi (\theta ){A_t},\operatorname{clip} (\psi (\theta ),1 - \varepsilon ,1 + \varepsilon ){A_t}} \right)} \right],
\label{eq_L_PPO}
\end{equation}
where $\varepsilon$ is the clipping parameter, and the $clip()$ operator constrains $\psi(\theta)$ to the interval $(1 - \varepsilon, 1 + \varepsilon)$ so as to limit undesirable policy deviations. The actor network updates its parameters $\theta$ by minimizing $L_{PPO}(\theta)$. Furthermore, a truncated generalized advantage estimation method is employed to compute $A_t$, expressed as \cite{schulman2017proximal}:
\begin{equation}
\begin{array}{c} 
\begin{array}{l} 
A_t = {\delta _t} + \gamma {\delta _{t + 1}} + \ldots + {\gamma ^{T - t + 1}}{\delta _{T - 1}} \\ 
where\quad \delta _t = {r_t} + \gamma {V_\varphi }\left( {{{\mathbf{s}}_{t + 1}}} \right) - {V_\varphi }\left( {{{\mathbf{s}}_t}} \right) \\ 
\end{array} 
\end{array},
\label{Eq_A_t}
\end{equation}
where $V_\varphi( {\mathbf{s}_t})$ represents the state-value function evaluated at $t$. The critic network weights are updated using the empirical estimates of $A_t$ obtained from each sampled batch. The corresponding loss for the critic is:
\begin{equation}
L_{critic}(\varphi ) = \frac{{\sum\limits_{t = 1}^T {\sqrt {{A_t}} } }}{T}.
\label{eq_L_critic}
\end{equation}
This loss formulation follows the standard practice adopted in PPO to train the critic network \cite{schulman2017proximal}, where the objective is to reduce the discrepancy between the estimated state values and their corresponding targets. Specifically, the incorporation of the advantage estimator $A_t$ supports the critic to accurately approximate $V_\varphi( {\mathbf{s}_t})$, which in turn provides reliable guidance for stabilizing the policy updates performed by the actor network. Such a design plays a critical role in reducing variance and improving convergence behavior in dynamic environments, similar to the one discussed in this paper. Readers seeking a more detailed explanation of the PPO framework are referred to the original work in \cite{schulman2017proximal}.

The pseudo-code of the PPO procedure proposed to address the problem \textbf{P1} is presented in Algorithm \ref{Alg:PPO_Algorithm} together with the structural depiction in Fig. \ref{PPO_architicture}. Lines 1--3 initialize all relevant agent components, such as the state representation, neural network parameters, and replay memory $\mathcal{D}$. In Lines 4--10, the agent interacts with the UAV environment in Fig. \ref{Fig1} by observing the current state, selecting actions, and receiving subsequent states and rewards, all of which are recorded in $\mathcal{D}$. Lines 11--16 describe the learning stage, during which the flying control agent samples mini-batches from the stored transitions to update and improve its policy ($\pi$). This cycle repeats until convergence to  optimal policy for solving \textbf{P1}.

\begin{algorithm}[t!]
\footnotesize
\begin{algorithmic}[1]
\renewcommand{\algorithmicrequire}{\textbf{Input:}}
\renewcommand{\algorithmicensure}{\textbf{Output:}}
\REQUIRE $N, R_{min}^{sec}$.
\ENSURE  Optimal $\mathbf Q, \boldsymbol p, \boldsymbol e^{h}, \boldsymbol{N}, \boldsymbol{S}, \boldsymbol{Q}$.
\STATE \textbf{Initialization}: Set $t=0$ and initialize the agent state $s_0$ using initial observations.
\STATE Randomly initialize the parameters of the actor \& critic networks $\theta$.
\STATE Initialize the experience memory buffer $\mathcal{D}$.
\FOR {episode = 1 to $T$}
    \STATE Reset the agent state $s(t)$.
    \FOR {$t = 1$ to $M$}
        \STATE Acquire the current state $s(t)$ from the environment using (\ref{Eq_stateSpace}).
        \STATE Select and execute action $a(t)$ based on (\ref{Eq_actionSpace}) following policy $\pi_{{\theta}_{old}}$.
        \STATE Obtain the reward using (\ref{Eq_rewardFunction}) and get subsequent state $s(t+1)$.
        \STATE Store tuple $\langle s(t), a(t), r(t), s(t+1) \rangle$ in $\mathcal{D}$.
        \IF {len(batch) $\geq$ batch size}
            \STATE Compute the advantage estimate $A_t$ according to (\ref{Eq_A_t}).
            \STATE Randomly sample a mini-batch of experiences from $\mathcal{D}$.
            \STATE Update actor and critic parameters $\theta$ to minimize $L_{PPO}(\theta)$ and $L_{critic}(\varphi)$ using (\ref{eq_L_PPO}) and (\ref{eq_L_critic}).
        \ENDIF
        \STATE Update parameters of target network;  $\theta_{old} \leftarrow \theta$.
    \ENDFOR
\ENDFOR
\end{algorithmic}
\caption{PPO-Based Joint Trajectory Optimization and Resource Allocation for UAV Networks}
\label{Alg:PPO_Algorithm}
\vspace{-1.mm}
\end{algorithm}

\vspace{-4.mm}
\subsection{Complexity of the Proposed PPO Algorithm}
The computational complexity of the proposed PPO algorithm, summarized in Algorithm \ref{Alg:PPO_Algorithm} and depicted in Fig. \ref{PPO_architicture}, can be assessed as follows \cite{Alwarafy2025Deep, shao2023secrecy}. PPO is generally executed within an Actor-Critic model \cite{schulman2017proximal}, and its computational cost primarily depends on the number of neural network operations carried out during each training episode. Because the actor and critic networks are constructed using fully connected layers, the complexity associated with a forward or backward pass is $\mathcal{O}\left(\sum_{k=0}^{K-1} C_k^{\mathrm{in}} C_{k+1}^{\mathrm{out}} \right)$, where $C_k^{\mathrm{in}}$ and $C_k^{\mathrm{out}}$ denote the input and output dimensions of the $k$th hidden layer, respectively, and $K$ is the number of hidden layers. In the considered design, the networks of the actor and critic share the same architecture, which results in a combined complexity of $\mathcal{O}\left(2\sum_{k=0}^{K-1} C_k^{\mathrm{in}} C_{k+1}^{\mathrm{out}} \right)$. Considering that Algorithm~\ref{Alg:PPO_Algorithm} iterates over $T$ training episodes and $M$ time slots per episode, the computational complexity of the PPO model can thus be formulated as $\mathcal{O}\left(2TM\sum_{k=0}^{K-1} C_k^{\mathrm{in}} C_{k+1}^{\mathrm{out}} \right)$.

\vspace{-0.4cm}
\section{Performance Evaluation}\label{Performance_Evaluation}
This section assesses the proposed PPO-driven DRL approach, described in Algorithm \ref{Alg:PPO_Algorithm}, for addressing the network optimization problem \textbf{P1}. First, the configuration of the simulated UAV-enabled network and the PPO agent architecture are specified. The numerical findings are then reported and discussed. In addition, an online-learning setting is examined to demonstrate the ability of the PPO agent to respond to time-varying and sudden network conditions, including changes caused by eavesdropper movement.

Our proposed method is evaluated against the following representative baselines:
\begin{itemize}
    \item The following policy-gradient DRL-based approaches:
    \begin{itemize}
        \item DDPG scheme, following the general approach in \cite{aboueleneen2023secure, Alwarafy2025Deep}; and
        \item Asynchronous Advantage Actor--Critic (A3C) algorithm, in line with \cite{Mahmoudi2025Beamforming,yang2021uav, Alwarafy2025Deep}.
    \end{itemize}
    
    \item A conventional genetic-algorithm (GA) benchmark, as considered in \cite{ciftler2021distributed,RN19,Alwarafy2024Rate, Alwarafy2025Deep}.
\end{itemize}
 
\vspace{-0.4cm}
\subsection{Simulation Setup} \label{sec4-a}
The UAV communication environment illustrated in Fig. \ref{Fig1} is adopted to evaluate the proposed framework. The network comprises 10--50 UAVs, which may operate in transmitting, jamming, or relaying modes, together with a single passive ground eavesdropper, all deployed within a circular area with a 200 m diameter. The transmitting and relaying UAVs forward their data to a tethered UAV hovering in the network, which serves as the flying center (F). The mobility and operating modes of all UAVs are governed by control actions generated by the learning agent, i.e., the tethered flying center, which centrally coordinates their movement and communication behavior. The remaining simulation parameters are selected in accordance with \cite{an2022secrecy,li2021secrecy,yan2024energy,liu2023joint,aboueleneen2023secure,Alwarafy2025Deep} and are summarized in Table \ref{Simulation_Setup}. Furthermore, the learning agent is implemented using the PPO DRL algorithm in Python with the PyTorch framework. All deep neural networks use hyperbolic tangent activation functions and batch-normalization layers to improve training stability, whereas the Adam optimizer updates the parameters of the actor and critic networks. The detailed hyperparameter settings of the learning PPO model are provided in Table \ref{Model_architicture}.

\begin{table}
    \caption{Simulation Parameters of Network}
    \centering
    \begin{tabular}{|c|c||c|c|}
        \hline
        \textbf{Notation} & \textbf{Value} & \textbf{Notation} & \textbf{Value} \\ \hline
        $N$ & 10--50 &   $N_e$ & 1  \\ \hline
        Bandwidth & 20 MHz &  $v_{max}$ & 4 m/s   \\ \hline
        $N_0, N_0^{'}$ & -174 dBm/Hz & $L$& 5 m  \\ \hline  
        $l_{i/r}$&[100, 200]Mbps  &   $n_{LoS}$  &2  \\ \hline 
        $h_{min}$& 20 m & $n_{N LoS}$  &3   \\ \hline 
        $h_{max}$& 80 m & $P^{max}$ & 30dBm  \\ \hline   
        $a$& $0.28$ & $\zeta_i$/$\zeta_j$ & 0.5\\ \hline 
        $\beta$&  $9.6$ &   $B^{max}$ &  500J \\ \hline
        $f_c$ & 2.4 GHz & $\lambda_{LoS}$& 1 dB \\ \hline 
        $\lambda_{NLoS}$& 20 dB & $d_{\text{min}}$& 10 m  \\ \hline  
    \end{tabular}
    \label{Simulation_Setup}
\vspace{-5.mm}
\end{table}

\begin{table}[t]
\centering
\caption{\textsc{PPO Model Hyperparameters}}
\label{Model_architicture}
\resizebox{\columnwidth}{!}{
\begin{tabular}{|l|c|}
\hline
\textbf{Parameter} & \textbf{Value}  \\ \hline
Learning rate & actor ($1\times10^{-3}$), critic ($8\times10^{-4}$)  \\ \hline
Batch size & $64$  \\ \hline
Horizon ($N$) & $20$  \\ \hline
No. epochs & $10$  \\ \hline
$\gamma$ & $0.99$  \\ \hline
GAE parameter $(\lambda)$ & $0.95$  \\ \hline
Hidden layers number & actor \& critic = $3$   \\ \hline
Neurons number & $(512, 512, 512)$ \\ \hline
Optimizer & Adam  \\ \hline 
Policy clip & $0.2$ \\ \hline
\end{tabular}%
}
\vspace{-3.mm}
\end{table}

\vspace{-0.45cm}
\subsection{Numerical Results} \label{Numerical Results}
This subsection reports the simulation results and examines the effectiveness of the proposed PPO-based solution relative to the considered benchmark schemes.

\begin{figure}[t!]
	\centering
	\includegraphics[width=7.cm, height=6cm]{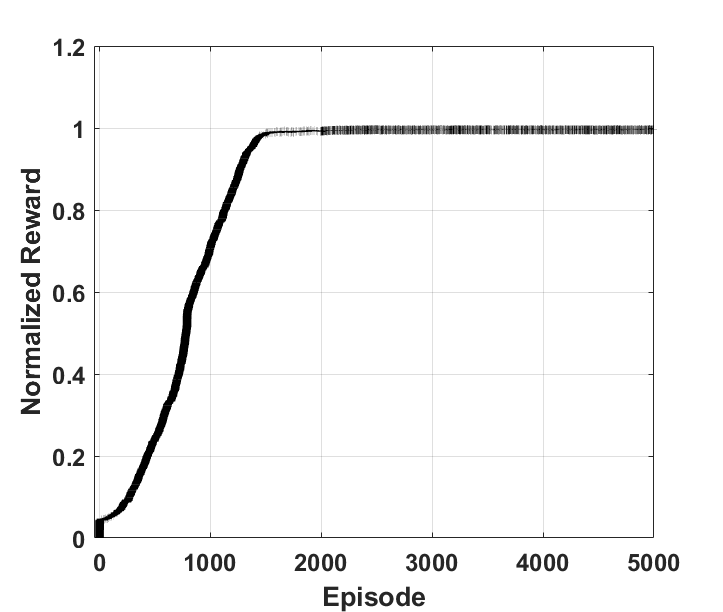}
	\caption{Convergence behavior of the proposed PPO agent. The reward curve converges around episode 1506 and stabilizes by episode 1992.}
	\label{reward}      
\vspace{-6.2mm}
\end{figure}

Fig. \ref{reward} shows the training reward of the proposed PPO algorithm. The increasing reward indicates that the agent learns an effective policy $\pi$ for solving optimization problem \textbf{P1}. The data show a successful learning run that largely converged around episode 1506 and remained highly stable afterward. Continuing training beyond roughly 1992 episodes delivered only marginal improvement.

\begin{figure}[t!]
	\centering
	\includegraphics[width=7.cm, height=6cm]{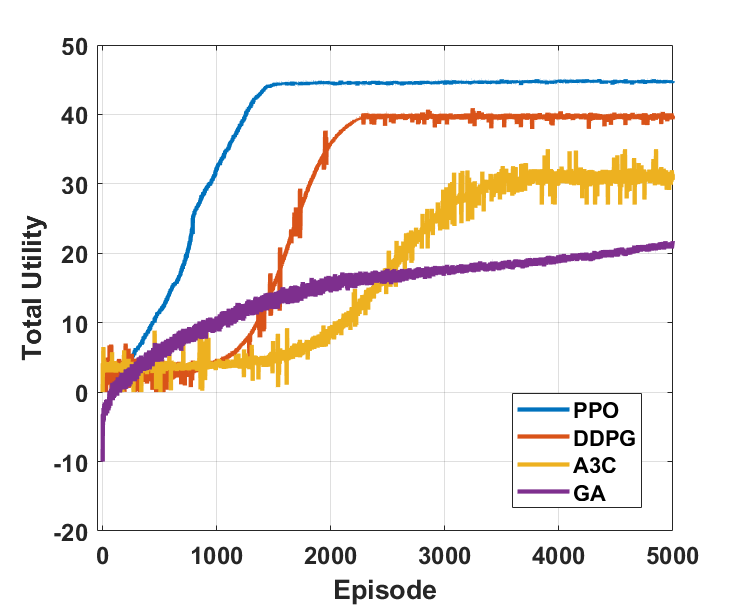}
	\caption{Total utility achieved by the proposed PPO algorithm and the benchmark solutions. PPO converges faster and more stably than the benchmark methods while attaining higher utility values.}
	\label{Utility}
\vspace{-6.mm}
\end{figure}

Fig. \ref{Utility} compares the total utility obtained by the proposed PPO algorithm with that achieved by the considered benchmark methods. Several important observations can be made from this figure. First, the PPO, DDPG, and A3C agents learn utility-maximizing policies for solving \textbf{P1} after approximately 1506, 2286, and 3498 episodes, respectively. However, PPO demonstrates the fastest and most stable convergence, as indicated by its smaller utility fluctuations, followed by DDPG and A3C. In contrast, the GA-based approach continues to improve throughout the evaluated horizon and does not converge within 5000 episodes. This behavior indicates its lower utility performance and weaker stability compared with the DRL-based approaches. Second, the proposed PPO algorithm achieves the highest total utility after convergence. Specifically, its converged utility exceeds those of DDPG, A3C, and GA by $5.06$, $13.73$, and $24.6$, respectively. These results confirm that our proposed PPO method provides a more efficient and stable solution to the considered optimization problem by achieving rapid policy learning and improved long-term utility.

\begin{figure*}[!t]
    \centering

    \subfloat[Normalized total secrecy rate\label{secrecy_rate}]{
        \includegraphics[width=0.3\textwidth]{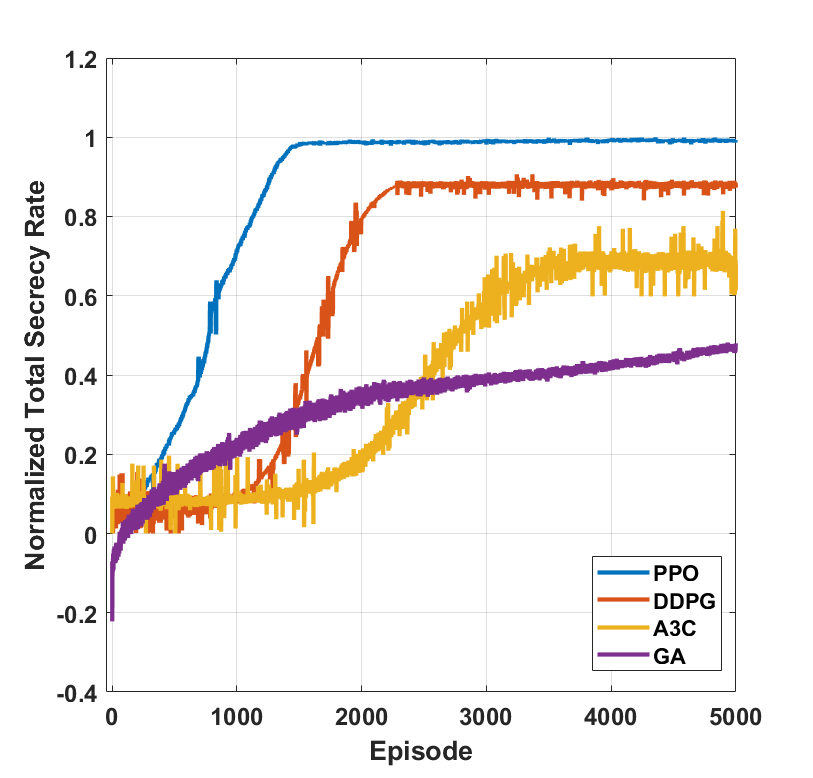}
    }
    \subfloat[Normalized total energy consumption\label{energy_effeciency}]{
        \includegraphics[width=0.3\textwidth]{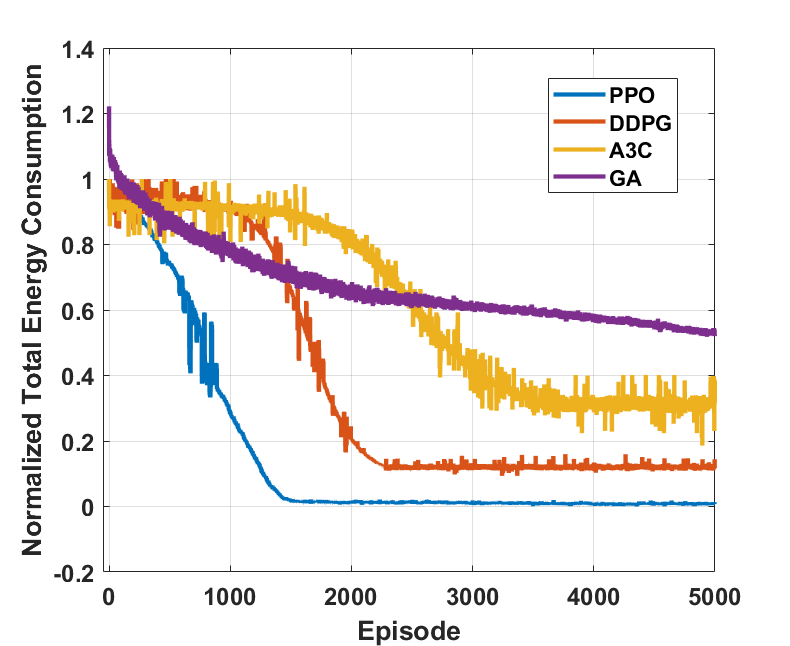}
    }
    \subfloat[Normalized total interference\label{interference}]{
        \includegraphics[width=0.3\textwidth]{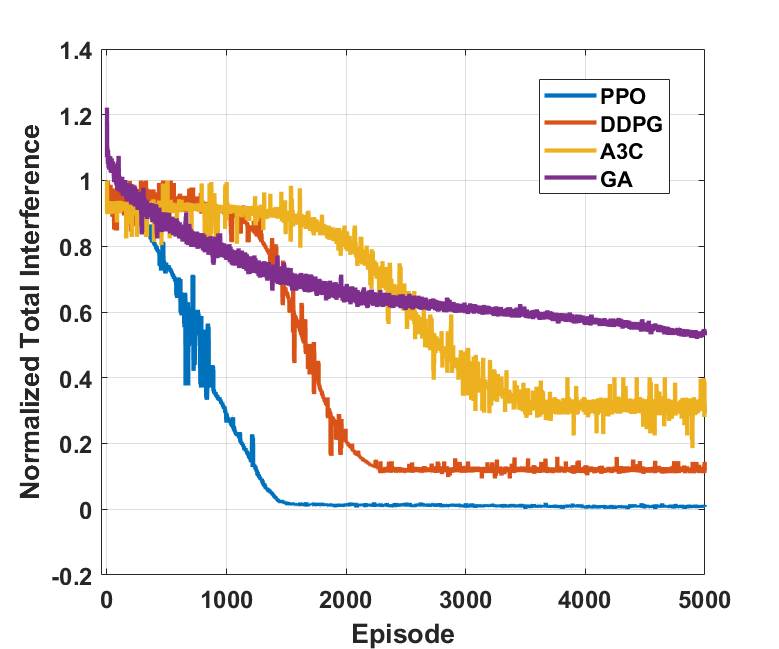}
    }

    \caption{Normalized performance comparison of the proposed PPO algorithm and the benchmark methods. PPO converges more stably and rapidly while attaining (a) the highest secrecy-rate performance, (b) the lowest energy consumption, and (c) the lowest interference level.}
    \label{secrecy_energy_interference}
\vspace{-7.mm}
\end{figure*}

Problem \textbf{P1} jointly seeks maximizing the total secrecy rate while minimizing the network energy consumption and interference. Figs. \ref{secrecy_rate}, \ref{energy_effeciency}, and \ref{interference} present the normalized total secrecy rate, total energy consumption, and total interference, respectively, achieved by our proposed PPO method and the benchmark schemes.

As depicted in Fig. \ref{secrecy_rate}, PPO attains the highest secrecy rate and reaches its stable operating region substantially earlier than the competing methods. The performance comparison is based on the average values over the last 1000 training episodes. Under this criterion, PPO improves the average secrecy rate by $12.76\%$, $44.37\%$, and $122.16\%$ relative to DDPG, A3C, and GA, respectively. This gain confirms that PPO learns a more effective resource-allocation and trajectory-control policy for secure UAV communications. The energy-consumption results in Fig. \ref{energy_effeciency} further demonstrate the advantage of the proposed method. PPO converges rapidly to a near-zero normalized energy consumption with limited residual variation. Compared with DDPG, A3C, and GA, it reduces the average energy consumption by $93.11\%$, $97.35\%$, and $98.5\%$, respectively. Similarly, Fig. \ref{interference} shows that PPO suppresses total network interference more efficiently and with greater stability than the benchmark algorithms. Specifically, PPO achieves average interference reductions of $93.29\%$, $97.43\%$, and $98.54\%$ compared with DDPG, A3C, and GA, respectively. Overall, the results verify that PPO provides a favorable balance between secrecy enhancement, energy conservation, and interference mitigation.

\begin{figure}[t!]
	\centering
	\includegraphics[width=7.5cm, height=5.8cm]{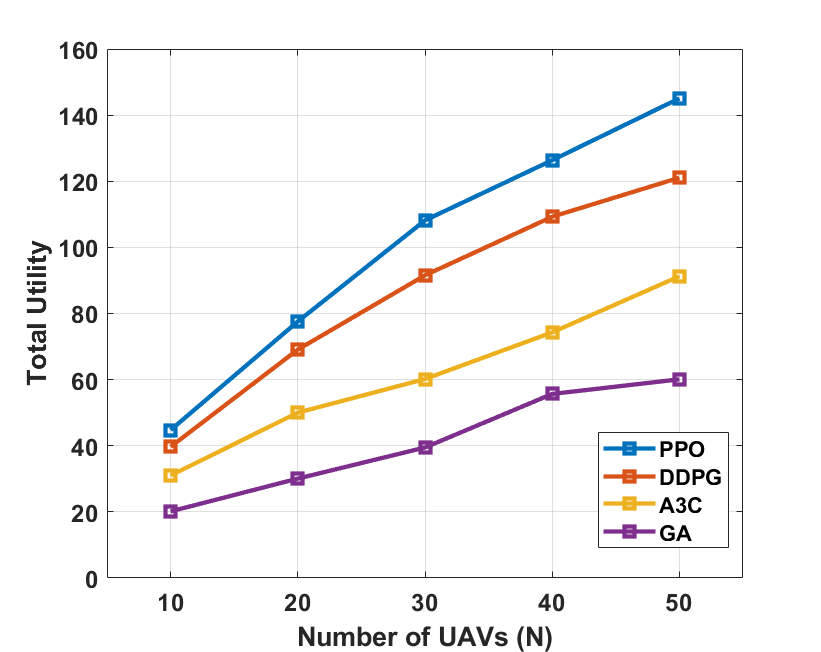}
	\caption{Total utility achieved by the proposed PPO algorithm and the benchmarks for different numbers of UAVs $N$. The PPO algorithm consistently achieves the highest utility and maintains an increasing performance advantage as the network size grows.}
	\label{Utility_vs_N}
\vspace{-5.mm}
\end{figure}

To examine the efficiency of our proposed PPO method in solving problem \textbf{P1} as the scale of the UAV network in Fig. \ref{Fig1} increases, Fig. \ref{Utility_vs_N} illustrates the total utility achieved by the proposed method and the benchmarks for different numbers of UAVs $N$. All methods achieve higher utility values as $N$ grows from 10 to 50, since a larger UAV network provides additional transmission, relaying, and jamming opportunities. However, the proposed PPO algorithm consistently delivers the highest utility across all network sizes, followed by DDPG, A3C, and GA.

More importantly, the performance advantage of PPO becomes increasingly pronounced as the network scale expands. For $N=10$, PPO exceeds the converged utilities of DDPG, A3C, and GA by $5.06$, $13.73$, and $24.6$, respectively. When $N$ increases to 50, these utility gaps widen to $23.99$, $53.81$, and $84.98$, respectively. This widening margin indicates that PPO scales more effectively with the size of the UAV network and can better exploit the additional degrees of freedom introduced by more UAVs. In particular, the result demonstrates the ability of PPO to maintain effective joint trajectory design, resource allocation, and role-selection decisions in larger and more complex network configurations.

\begin{figure}[t!]
	\centering
	\includegraphics[width=8.5cm, trim=0 12mm 0 0, clip]{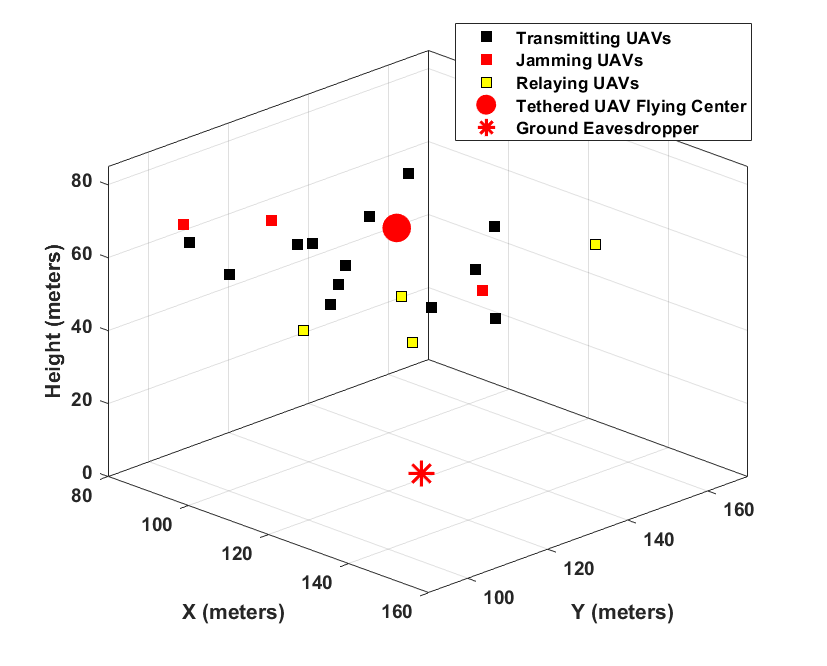}
	\caption{Optimized 3D locations and role deployment for $N=20$, including transmitting, jamming, and relaying UAVs, the tethered UAV flying center, and the ground eavesdropper.}
	\label{uav_role_deployment_n20}
\vspace{-7.mm}
\end{figure}

To illustrate the solution of problem \textbf{P1} at the UAV-network level, Fig. \ref{uav_role_deployment_n20} presents the optimized 3D role deployment for the case of $N=20$, whose corresponding total utility is reported in Fig. \ref{Utility_vs_N}. The proposed PPO framework jointly determines the role-assignment variables $\boldsymbol{N}$ and $\boldsymbol{S}$, as well as the spatial positions of the UAVs $\mathbf{Q}$ and the tethered UAV flying center $\boldsymbol{q}_{F}$. As shown in Fig. \ref{uav_role_deployment_n20}, the optimized deployment consists of $N_t=13$ transmitting UAVs, $N_j=3$ jamming UAVs, and $N_r=4$ relaying UAVs (i.e., $\boldsymbol{N}=[ 13, 3, 4]$ in constraint $C_{11}$ in problem \textbf{P1}). The transmitting UAVs provide the intended communication links, whereas the jamming and relaying UAVs are selectively positioned to enhance physical-layer security and connectivity, respectively. In addition, the optimized location of the tethered UAV flying center $\boldsymbol{q}_{F}$ is shown relative to the UAVs deployment and the ground eavesdropper. This role-aware 3D configuration demonstrates the ability of the proposed PPO algorithm to jointly coordinate UAV functions and locations to maximize the network utility.

\vspace{-0.35cm}
\subsection{Online Deployment of the Proposed PPO Model} \label{model_Adaptivity}
\begin{figure}[t!]
	\centering
	\includegraphics[width=7.5cm, trim=0 12mm 0 0, clip]{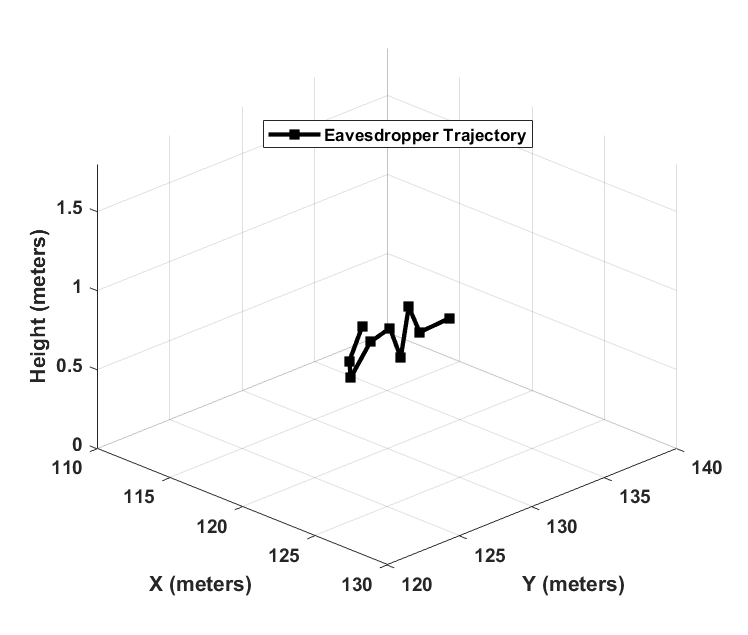}
	\caption{3D trajectory of the ground eavesdropper generated according to the Random Waypoint Mobility Model.}
	\label{Mobility_pattern}
\vspace{-5.mm}
\end{figure}

To test the practical robustness of the proposed PPO framework, we investigate its ability to adapt online to changes in the ground eavesdropper's location under dynamic network conditions. Specifically, the ground eavesdropper follows the Random Waypoint Mobility Model described in Subsection II.E similar to \cite{Zhong2023On,Alwarafy2024Rate,Alwarafy2025Deep}, which causes all UAVs channel gains to update accordingly. After the initial convergence reported in Subsection \ref{Numerical Results}, the eavesdropper updates its position every 1000 episodes. Fig. \ref{Mobility_pattern} depicts the resulting eavesdropper trajectory within the considered deployment area for $N=10$.

\begin{figure}[t!]
	\centering
	\includegraphics[width=7.cm, height=5.8cm]{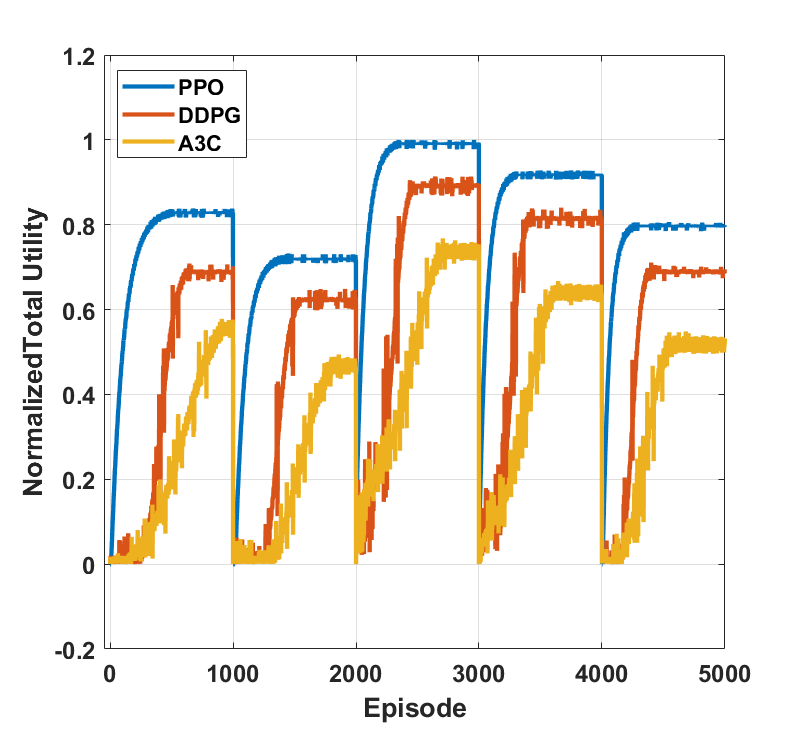}
	\caption{Online adaptation performance under eavesdropper mobility. The eavesdropper updates its location every 1000 episodes, and the proposed PPO agent efficiently relearns an effective policy after each network change. PPO achieves faster convergence and higher normalized utility than the DDPG and A3C benchmark algorithms.}
	\label{Utility_mobility}
\vspace{-5.mm}
\end{figure}

Based on the mobility scenario shown in Fig. \ref{Mobility_pattern}, Fig. \ref{Utility_mobility} shows the normalized total utility, which represents the objective value of problem \textbf{P1}, for the proposed PPO algorithm and the DDPG and A3C benchmark methods during online deployment. The GA method is not included because it does not converge to a feasible solution within the considered 1000-episode adaptation period. Three observations can be made from Fig.~\ref{Utility_mobility}. First, PPO quickly adapts to changes in the eavesdropper location and learns a new effective policy after each change. For example, during episodes 2000--3000, PPO converges after 357 episodes, while DDPG and A3C require 467 and 701 episodes, respectively. Second, all DRL algorithms adapt faster over time. When the first adaptation window (episodes 0--1000) is compared with the final window (episodes 4000--5000), the convergence speed improves by about $1.9$, $1.7$, and $1.7$ times for PPO, DDPG, and A3C, respectively. This result shows that the agents use their previous learning experience to respond more efficiently to later network changes. Finally, PPO achieves the highest normalized utility in all adaptation windows. Overall, these results confirm that our proposed PPO method provides faster adaptation, better stability, and higher utility when the tethered-UAV network experiences abrupt environmental changes.

\vspace{-0.4cm}
\section{Conclusion}\label{Conculusion}
In this work, we developed a DRL framework based on the PPO algorithm to solve a challenging multi-objective optimization problem in UAV-enabled wireless networks with energy harvesting and adversarial eavesdropping. The problem was formulated as a combinatorial non-convex optimization task that jointly optimizes the secrecy rate, energy consumption, and interference by controlling UAV 3D mobility, transmit, jamming, and relaying power, harvested energy, UAVs' dynamic role assignment, and the trajectory of a tethered UAV flying center. Due to the high dimensionality, mixed-integer decision space, and strong coupling among system variables, conventional optimization techniques become computationally prohibitive. To overcome these challenges, a PPO-based DRL approach was employed to efficiently learn adaptive control policies that capture the network’s complex dynamics. Extensive simulation results confirm that the proposed algorithm converges rapidly and consistently which outperforms benchmark schemes, including DDPG-, A3C-, and GA-based solutions, in terms of overall utility and stability. As a promising direction for future research, the proposed framework can be extended to scenarios involving the coexistence of both flying and ground eavesdroppers.

\vspace{-5.mm}
\bibliographystyle{IEEEtran}
\bibliography{References}
\end{document}